# On the validity of the discrete ordinates method in spherical geometry

Charles H. Aboughantous

*Louisiana State University, Department of Physics and Astronomy, Baton Rouge, LA 70803*

**Abstract.** The discrete ordinates method with angular parameters is discussed in light of the recent extension of its underlying hypothesis. The analysis shows that the method, although it compares very well with the trajectory of photons model, it addresses a boundary value problem different from the problem of flow of energy in the interior of spheres.

**Introduction**

The discrete ordinates method in spherical geometry with central symmetry was revisited recently with a new approach that retains the structure of the original boundary value problem in two respects: the angular parameters and its consequential starter specific intensity [1]. The angular parameters approximate the angular derivative and they are constrained to the condition of constant specific intensity* in the asymptotic region. The starter intensity is needed to initiate the solution of $N$ equations in $N+1$ unknowns that are generated with the angular parameters [2]. In what follow, we examine the validity of this discrete ordinates structure particularly in light of the hypothesis adopted for the development in [1].

**The divergence model of transfer**

The new approach relaxes the asymptotic condition of the angular parameters. It is sufficient to assume an isotropic specific intensity at the center of the sphere. This assumption was justified by observing that if we multiply the equation:

$$\left\{\mu\partial_r\psi + \frac{\eta^2}{r}\partial_\mu\psi\right\} + \sigma\psi = Q \qquad (1)$$

---
\* The angular neutron flux of the original paper is interpreted here as specific intensity and scalar flux as total intensity.

by $r$, then, according to the argument in [1], all the terms of the equation vanish at the center. Here $\partial_r$ is the tensor symbol of the derivative with respect to $r$, $\eta = (1 - \mu^2)^{1/2}$, $\psi$ the specific intensity, $\sigma$ opacity, $Q$ emission density, and the braces quantity is the divergence. The result of this multiplication, according to [1], is an isotropic intensity at the center of the sphere inferred from $\eta^2\partial_\mu\psi_{r=0} = 0$.

The divergence term is singular at $r = 0$. It yields an intensity in $1/r^2$ its spatial derivative varies as $1/r^3$, whether the sphere is vacuum or any other material, with or without a source. Therefore, whether $rQ$ vanishes at the center or not, neither $r\psi$ nor $r\partial_r\psi$ vanish at $r = 0$, which invalidates the argument for the assumption of an isotropic intensity at the center.

We realize that eq. (1) is the linearized form of the transfer equation, that is, it is not valid at $r = 0$. It is difficult to justify the hypothesis of isotropic specific intensity at the center. On the other hand, if we legitimize the multiplication by $r$ in the manner suggested in [1], the resulting isotropic intensity at the center is only one contribution to the physical intensity. The contributions from absorption and source emission are yet to be accounted for in order to obtain the integral intensity at the center.

Equation (1) defines a boundary value problem completely specified by one natural boundary condition at the surface: the boundary intensity $\psi_R$, where $R$ is the radius of the sphere. If we impose an additional boundary condition descriptive of the nature of the intensity at the center, such as the isotropic intensity suggested in [1], we will be solving a boundary value problem of our own which may not necessarily be representative of the actual flow of energy in the sphere. It was shown that the spatial derivative is always negative and $\psi$ is function of $\mu$ at $r \to 0$, that is, the solution is anisotropic at the center [3]. Indeed, the coefficient $\mu$ of $\partial_r\psi$ guarantees that $\psi$ is always anisotropic at the center, except in vacuum.

It was shown that if the specific intensity is needed only for calculating the energy stored in the sphere and the energy flow at the surface, the singularity at the center can be bypassed. This can be done by shifting the spatial variable so that the new variable is $\vartheta = a + r$, where $a$ is any positive number smaller, equal to or larger than the radius $R$, and $r$ varies on $[0, R]$. The integral over the volume of the now $\vartheta$-sphere produces the exact energy stored in the $R$-sphere under the conditions of validity of the linearized transfer equation [3].

The other shortcoming of the discrete ordinates of [1] is the starter intensity. It is inherited from the original formulation of the discrete ordinates. This is in effect a redundant boundary condition on the angular domain; the angular boundary condition is already contained in the boundary intensity $\psi_R$.

The starter intensity is taken to be the solution of the transfer equation in slab geometry by simply setting $\mu = -1 \Leftrightarrow \eta = 0$ in eq. (1). This practice is not appropriate for this class of equations. The angular derivative of eq. (1) couples the intensities in different directions asymmetrically in the sense that the integral of $\partial_\mu \psi$ does not vanish on $[-1, +1]$. If we remove it from the equation then we will be solving a boundary value problem different from the one defined by the original equation. A good approximation for the starter intensity should be consistent with $1/r^2$ intensity. It could be the case that plane geometry starter intensity is a reasonable approximation in a large sphere for $r \sim R$. It is unlikely to be the case in the interior of the sphere and certainly not in the proximity of the center. It was shown that it is possible to construct a set of discrete ordinates that closes the equations without necessitating starter intensity if un-normalized circular functions are used [3].

Now we address the question, can we retain the angular parameters of the discrete ordinates and start the solution with intensity in $1/r^2$, assuming we have a reasonable approximate prescription for that intensity? The short answer in general is no. The angular parameters are evaluated from a recursion relation conditioned by a constant asymptotic intensity [2]. The $1/r^2$ intensity is not constant anywhere in a finite sphere and therefore it is not consistent with the structure of the angular parameters. The same applies if the intensity is assumed isotropic at the center.

**The trajectory representation of transfer**

It is instructive at this point to examine the problem in a different representation. Consider a sphere of radius $R$ with isotropic and uniformly distributed boundary intensity $\psi_R$. Assume further that the sphere is vacuum; the expressions are simple and equally instructive as in any other sphere. If we convert the divergence into an ordinary derivative along the trajectory of a photon as it is done in [1], then eq. (1) becomes $\partial_s \psi = 0$ and its solution is $\psi_s = \psi_R$. At first sight, this identity suggests as if the total intensity is constant in $r$. Not necessarily, the total intensity is obtained as a line integral on an arc on the surface of the sphere. Therefore, whether $\psi_s$ is constant or not, the limits of integration must be explicit in $r$. This is the case in earnest in the medium exterior to a sphere its radiance is $\psi_R$. Then we will have:

$$\varphi_r \propto \int_{\mu_0}^{+1} \psi_s \, d\mu \quad r \geq R \tag{2}$$

where $\mu_0 = (1 - R^2/r^2)^{1/2}$ [1]. If the medium is vacuum and the radiance is isotropic and uniformly distributed, then:

$$\varphi_r \propto \psi_R \left[ 1 - \left(1 - R^2/r^2\right)^{1/2} \right] \tag{3}$$

This solution and the one obtained with discrete ordinates of [3] are numerically congruent for $\forall r \geq R$. This result shows that the trajectory model is compatible with the formal divergence representation in the exterior of a radiating sphere. The two representations reproduce the same physical quantity, the total intensity $\varphi$. It is not so in the interior of the sphere. There we would have:

$$\varphi_r \propto \int_{-1}^{+1} \psi_s \, d\mu \quad r \leq R \tag{4}$$

That is, the total intensity is uniform in the interior of our sphere. A closer look at this result reveals that eq. (4) is a statement of self-denial.

Indeed, if the intensity is uniform in the sphere, then its value is the same at the center and at the surface. By recognizing that the intensity at the center is the contribution from the intensity emitted normally to the surface, let $\psi_R$ be the normal intensity. Then the total normal energy is $4\pi R^2 \psi_R = 4\pi\varepsilon^2 \psi_\varepsilon$ where $\varepsilon$ is the radius of a concentric sphere. Clearly, the normal intensity $\psi_{\varepsilon\to 0}$ is different from $\psi_R$, which denies the validity of eq. (4).

It appears that the intensity $\psi_s$ is a kernel for the general solution for eq. (1). The line integral of the kernel yields the general solution for the total intensity if at least one limit of integration is explicit in $r$. This happened to be the case in the exterior but not in the interior of the sphere.

**Conclusion**

We conclude from the forgoing that the trajectory model is not a valid boundary value problem of radiative transfer in the interior of a sphere. Consequently, a discrete ordinates solution that is designed to compare with the solution in the trajectory model in the interior of a sphere represents a boundary value problem different from the problem of flow of energy in spheres. This appears to be the case of the discrete ordinates method of [1].